\documentclass[aps,pra,amsmath,twocolumn,amssymb,floatfix,showpacs,superscriptaddress,nofootinbib,longbibliography]{revtex4-1}
\usepackage{MnSymbol}
\usepackage{braket}
\usepackage[dvipsnames]{xcolor}
\usepackage{float}
\usepackage{subfigure}
\usepackage{tikz}
\usepackage[colorlinks=true,linktoc=page,linkcolor=MidnightBlue,citecolor=purple,urlcolor=violet]{hyperref}
\usepackage{multirow}

\newcommand{\vect}[1]{\boldsymbol{\mathrm{#1}}}

\mathchardef\mhyphen="2D 

\newcommand{\ie}{{i.e.,\,\,}}

\newcommand\bea{\begin{eqnarray}}
	\newcommand\eea{\end{eqnarray}}
\newcommand\beq{\begin{equation}}  
	\newcommand\eeq{\end{equation}}

\newcommand{\non}{\nonumber}  
\usepackage[normalem]{ulem}
\definecolor{lime}{HTML}{A6CE39}
\usepackage{sidecap,tikz}
\DeclareRobustCommand{\orcidicon}{\hspace{-1.0mm}
	\begin{tikzpicture}
		\draw[lime, fill=lime] (0.0,0.0) 
		circle [radius=0.15] 
		node[white] {{\fontfamily{qag}\selectfont \tiny \,ID}};
		\draw[white, fill=white] (-0.0525,0.095) 
		circle [radius=0.007];
	\end{tikzpicture}
	\hspace{-3.0mm}
}
\foreach \x in {A, ..., Z}{\expandafter\xdef\csname orcid\x\endcsname{\noexpand\href{https://orcid.org/\csname orcidauthor\x\endcsname}{\noexpand\orcidicon}}
}

\allowdisplaybreaks

\begin{document}

\title{Floquet second-order topological Anderson insulator hosting corner localized modes}  

\author{Arnob Kumar Ghosh\orcidA{}}
\email{arnob.ghosh@physics.uu.se}
\affiliation{Institute of Physics, Sachivalaya Marg, Bhubaneswar-751005, India}
\affiliation{Homi Bhabha National Institute, Training School Complex, Anushakti Nagar, Mumbai 400094, India}
\affiliation{Department of Physics and Astronomy, Uppsala University, Box 516, 75120 Uppsala, Sweden}

\author{Tanay Nag\orcidB{}}
\email{tanay.nag@hyderabad.bits-pilani.ac.in}
\affiliation{Department of Physics, BITS Pilani-Hyderabad Campus, Telangana 500078, India}

\author{Arijit Saha\orcidC{}}
\email{arijit@iopb.res.in}
\affiliation{Institute of Physics, Sachivalaya Marg, Bhubaneswar-751005, India}
\affiliation{Homi Bhabha National Institute, Training School Complex, Anushakti Nagar, Mumbai 400094, India}

\begin{abstract}
The presence of random disorder in a metallic system accounts for the localization of extended states in general. On the contrary, the presence of disorder can induce topological phases hosting metallic boundary states out of a non-topological system, giving birth to the topological Anderson insulator phase. In this context, we theoretically investigate the generation of an out of equilibrium higher-order topological Anderson phase in the presence of disorder potential in a time-periodic dynamical background. In particular, the time-dependent drive and the disorder potential concomitantly render the generation of Floquet higher-order topological Anderson insulator~(FHOTAI) phase, while the clean, undriven system is topologically trivial. We showcase the generation of FHOTAI hosting both $0$- and $\pi$-modes. Most importantly, we develop the real space topological invariant- a winding number based on chiral symmetry to characterize the Floquet $0$- and $\pi$-modes distinctly. This chiral winding number serves the purpose of the indicator for the topological phase transition in the presence of drive as well as disorder and appropriately characterizes the FHOTAI.
\end{abstract}

\maketitle

\section{Introduction} \label{introduction}
Disorder is an integral part of condensed matter systems, which may originate via crystal defects, impurity atoms, external perturbations, etc., and is almost inescapable~\cite{VojtaDisorder2019}. When the disorder strength is adequate, it can translate a metallic state to a localized insulating state~\cite{AnsersonPR1958}.
In the case of a topological insulator (TI), the boundary states are immune to weak and symmetry-preserving disorder since backscattering is forbidden for such states~\cite{kane2005quantum,BHZPRL2006,hasan2010colloquium,qi2011topological}. However, it has been theoretically proposed~\cite{JianLiPRL2009,GrothPRL2009,XingYPRB2011,SongPRB2014,ShapourianPRB2016,YuNatureRev2021,VeluryPRB2021,MadeiraPRB2022,TangPRA2022,LapierrePRL2022,DinhDuyPRB2022} and later experimentally demonstrated~\cite{EricScience2018,Stutzer2018,ZangenehAdvMater2020} that a system can host non-trivial phase in the presence of disorder, which otherwise is non-topological in the clean case. These disorder-induced TIs are coined as topological Anderson insulators~(TAIs).

Recently, higher-order TIs~(HOTIs) hosting $(d-n)$-dimensional boundary modes, have acquired significant research interest~\cite{benalcazar2017,benalcazarprb2017,Song2017,Langbehn2017,schindler2018,Franca2018,wang2018higher,Geier2018,Khalaf2018,ezawa2019second,luo2019higher,Roy2019,RoyGHOTI2019,Trifunovic2019,agarwala2019higher,Szumniak2020,Ni2020,BiyeXie2021,trifunovic2021higher}; with $d$ and $n$ being the dimension and topological order of the system, respectively such that $n\ge 2$. The disorder-induced topological phase in the context of HOTIs has also been investigated where a symmetry-preserving disorder configuration is employed to generate higher-order TAI~(HOTAI)~\cite{LiPRL2020,arakiPRB2019,WangPRRdisorder2020,ybyangPRB2021,jhwangPRL2021,yshuPRB2021,ZhangExptPRL2021,ZhangHOTAIPRB2021,PengQuasiPRB2021,LiuPRB2021,LiPRBL2022,WuPRAL2022}. The latter has also been demonstrated experimentally employing a electrical circuit setup~\cite{ZhangExptPRL2021}. In recent times, HOTIs have been generated in a driven out-of-equilibrium scenario while starting from a trivial static phase~\cite{Bomantara2019, Nag19,YangPRL2019,Seshadri2019,Martin2019,Ghosh2020,Huang2020,HuPRL2020,YangPRR2020,ZhangYang2020,chaudharyphononinduced2020,GongPRBL2021,Nag2021,JiabinYu2021,Vu2021,ghosh2021systematic,du2021weyl,WuPRBL2021}. These dynamical HOTIs, so-called Floquet HOTIs~(FHOTIs), can possess both $0$- as well as $\pi$-modes while the latter do not have any static analog~\cite{Martin2019,Huang2020,HuPRL2020,GongPRBL2021,JiabinYu2021,Vu2021,ghosh2021systematic}. 

In the context of periodically driven systems, disorder can foster the generation of Floquet TAIs~(FTAIs)~\cite{TitumPRL2015,RoyDisorderPRB2016,TitumPRX2016,TitumPRB2017,ChenPRB2018,DuPRB2018,MenaPRB2019,NingHallPRB2022,BhargavaPRB2022}. The FTAIs host only $0$-modes in the high-frequency limit and can still be characterized by the static topological invariants in real space~\cite{TitumPRL2015}. However, when the driving frequency 
is comparable to the system's bandwidth, the system may possess $\pi$-modes along with concurrent $0$-modes~\cite{Kitagawacharacterization2010,JiangColdAtomPRL2011,Rudner2013,Piskunow2014,Usaj2014,Yan2017,Eckardt2017,NHLindner2020}. In this scenario, one needs to construct a real space topological invariant than can capture the topological phases of the system unambiguously~\cite{TitumPRX2016,RoyDisorderPRB2016}. 

Given this framework, the HOTAI has been investigated in a driven setup employing the high-frequency approximation~\cite{NingPRB2022}. This setup only exhibits higher-order modes around quasienergy zero. However, it is not yet known in the current literature how to engineer a Floquet HOTAI~(FHOTAI) hosting both the $0$- and anomalous $\pi$-modes. Moreover, the clean FHOTI hosting $0$ and $\pi$-modes can be characterized employing quadrupolar and octupolar motions~\cite{Huang2020,GhoshDynamical2022}. These topological invariants are based on momentum space formalism and require translational invariance and mirror symmetries. Thus, it is essential to develop a real space topological invariant that can be employed to characterize the $0$- and $\pi$-modes distinctly in the out-of-equilibrium FHOTAI phase while explicitly not depending upon the mirror symmetry protection.

In this article, we consider a step-drive protocol in the presence of disorder to engineer the novel Floquet second-order TAI~(FSOTAI) phase, which doesn't have any clean and static analog. We develop a real space topological index that can successfully characterize the $0$ and $\pi$ modes for a chiral-symmetry-protected FSOTAI. 

The remainder of this manuscript is organized as follows. In Sec.~\ref{SecII}, we introduce our driving protocol and demonstrate the generation of FSOTAI. Sec.~\ref{SecIII} is devoted 
to the discussion of topological characterization, where we introduce and develope the topological invariant that we employ to characterize the FSOTAI phase. Finally, we conclude our article with a brief summary and discussion in Sec.~\ref{SecIV}. In Appendix~\ref{App:cleanFHOTI}, we demonstrate the quasienergy spectra of clean FHOTI. In Appendix~\ref{App:LSSP}, we discuss the localization properties of the eigenstates of the disordered Floquet operator. In Appendices~\ref{App:FOchiral} and \ref{App:endstates}, we discuss how the Floquet operator and the corner modes transform under chiral symmetry, respectively. We demonstrate a harmonically driven disordered setup in Appendix~\ref{App:Harmonicdrive} and also exhibit that our formalism works well for a different model discussed in Appendix~\ref{App:BBHmodelsec}.

\section{Driving protocol and generation of FSOTAI} \label{SecII}
We consider a three-step driving protocol based on a quantum spin Hall insulator with a fourfold rotation ($C_4$) and time-reversal ($\mathcal{T}$) symmetry breaking Wilson-Dirac mass term as~\cite{schindler2018,agarwala2019higher,Nag19,Ghosh2020,ghosh2021systematic}
\begin{align}
H (t)&=  h_1 \  ;   \qquad \qquad t \in [0, T/4] \ , \nonumber \\
&= h_2 \  ; \qquad \qquad t \in (T/4,3T/4] \  , \nonumber \\
&= h_3 \  ; \qquad \qquad t \in (3T/4,T ] \ .
\label{drive}
\end{align}
Here, $T$ represents the time period of the drive, and the step Hamiltonians $h_{1,2,3}$ based on a square lattice geometry reads as
\begin{align}
    h_1&=h_3= \sum_{i,j} c_{i,j}^\dagger \left( J_1 + V_{ij} \right) \Gamma_1 c_{i,j} \ , \non  \\
    h_2&= \sum_{i,j} c_{i,j}^\dagger \frac{J_2}{2} \big( \Gamma_1  c_{i+1,j}+ \Gamma_1  c_{i,j+1}  + i \Gamma_2 c_{i+1,j} + i \Gamma_3 c_{i,j+1} \non \\
    & \qquad \qquad +\Gamma_4  c_{i+1,j}- \Gamma_4  c_{i,j+1} \big) + {\rm h.c.} \ ,
\end{align}
where $c_{i,j}$ represent electron creation operator at lattice site $(i,j)$ with $i$ and $j$ denoting the lattice sites along $x$- and $y$-directions, respectively. Here, $J_1$ and $J_2$ symbolize the driving strengths, and $V_{ij}$ represents the uniformly distributed onsite random disorder such that $V_{ij} \in \left[ -w/2,w/2 \right]$. The $4 \times 4$ $\vect{\Gamma}$ matrices are given as $\Gamma_1=\sigma_z s_0$, $\Gamma_2=\sigma_x s_z$, $\Gamma_3=\sigma_y s_0$, $\Gamma_4=\sigma_x s_x$; with $\vect{\sigma}$ and $\vect{s}$ acting on the orbital and spin degrees of freedom, respectively. Here, $h_2$ breaks the mirror symmetries. However, $h_{1,2,3}$ respects the chiral symmetry $S=\sigma_x s_y$. The chiral symmetry plays a pivotal role in defining the topological invariant for this driven system. Here, the step Hamiltonians $h_{1,3}$ and $h_2$ represent a trivial band insulator and second-order TI, respectively. Note that, we add onsite random disorder only in the step Hamiltonian $h_{1,3}$, while we turn off the disorder in the second step, \ie in $h_2$. Similar type of quenched disorder in terms of random magnetic field has been previously employed in 
case of many-body localization in a periodically driven spin system~\cite{PonteMBLPRL2015}. We here adopt the similar path for a fermionic system in terms of quenched random onsite disorder.

\begin{figure}
	\centering
	\subfigure{\includegraphics[width=0.4\textwidth]{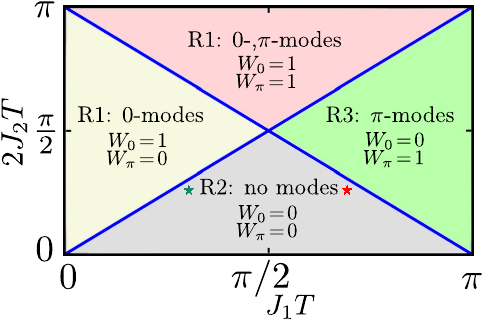}}
	\caption{We illustrate the phase diagram in the $J_1T \mhyphen J_2 T$ plane considering the driving protocol introduced in Eq.~(\ref{drive}) for $V_{ij}=0$ \ie a clean driven system. The phase diagram is divided into four parts- R1, R2, R3, and R4.
	}
	\label{CleanPhase}
\end{figure}

To solve this time-dependent problem, we construct the time-evolution operator employing the time-ordered~(TO) notation as
\begin{align}
    U(t,0)= {\rm TO} \exp \left[ -i \int_0^t  H(t') dt' \right] \ .
\end{align}
The Floquet operator $U(T,0)$ reflects the outcomes of the driven system. In the absence of disorder \ie $V_{ij}=0$, one can construct the Floquet operator in momentum space to obtain the topological phase diagram analytically~\cite{Huang2020,ghosh2021systematic,GhoshDynamical2022}. In particular, one obtains the quasienergy spectra as $U(T,0;\vect{k}) \ket{\Psi(\vect{k})}=E(\vect{k})\ket{\Psi(\vect{k})}$. The corresponding phase diagram is depicted in Fig.~\ref{CleanPhase} by considering $E(\bar{\vect{k}})=0/ \pm \pi$; with $\bar{\vect{k}}=(0,0)/(\pi,\pi)$~\cite{ghosh2021systematic,GhoshDynamical2022,GhoshTimedynamics2023}. The blue lines represent the phase boundaries where the bulk gap closes at quasienergies $0$ or $\pi$. To understand different parts of this phase diagram, one can diagonalize the corresponding Floquet operator considering a finite size system with open boundary condition~(OBC) along both directions. We observe that the phase diagram is divided into four parts- region 1~(R1) supporting only $0$-modes, region 2~(R2) exhibiting a trivial phase, region 3~(R3) with only $\pi$-modes, and region 4~(R4) hosting both $0$- and $\pi$-modes. In Appendix~\ref{App:cleanFHOTI}, we demonstrate the quasienergy spectra in these different topological regime explicitly for a clean driven system.

\begin{figure}
	\centering
	\subfigure{\includegraphics[width=0.49\textwidth]{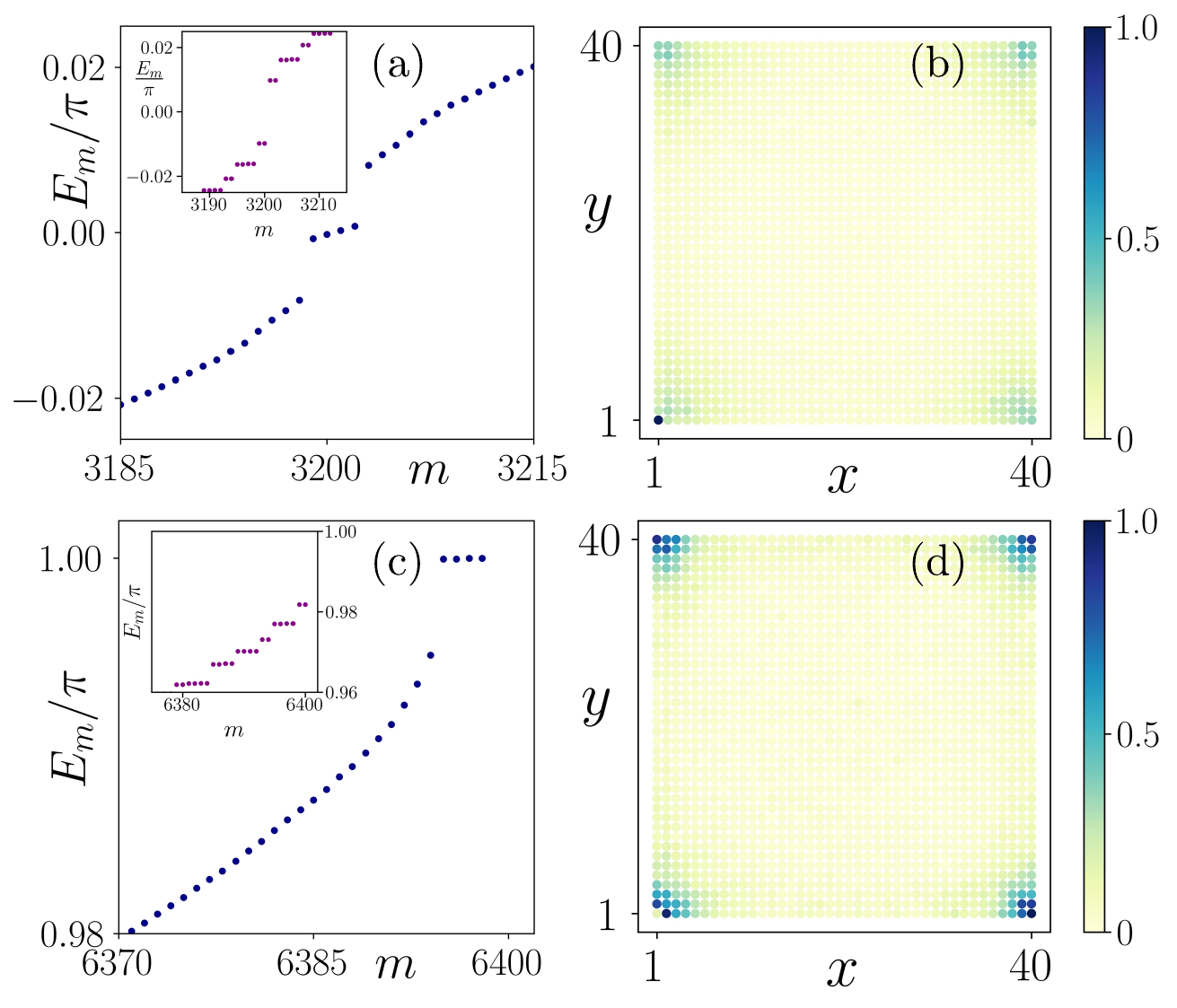}}
	\caption{In panels (a) and (b) [(c) and (d)], we depict the eigenvalue spectrum $E_m$ close to quasienergy $0$ [$\pi$] as a function of the state index $m$ and the LDOS at $E_m=0~[\pi]$ as a function of the system dimensions. While the absence of any topological modes around $E_m=0~[\pi]$ is evident from the insets of (a)~[(c)] where $w=0$. The parameters are chosen in regime R2 of Fig.~\ref{CleanPhase}. In particular, we choose $(J_1,J_2)=(0.8,\pi/8)$ (highlighted by a green star in Fig.~\ref{CleanPhase}) and $w=1.0$ [$(J_1,J_2)=(2.31,\pi/8)$ (indicated by a red star in Fig.~\ref{CleanPhase}) and $w=1.5$] for panels (a) and (b) [(c) and (d)]. We consider $T=2.0$ and our system comprises of $40 \times 40$ lattice sites with $100$ random disorder configurations.
	}
	\label{Eigenvalue}
\end{figure}

Having investigated the phases for the driven clean system, we discuss the generation of FSOTAI in 
the presence of disorder while starting from a clean topologically trivial phase. First, we choose a point in the phase space, indicated by a green star in Fig.~\ref{drive} (within R2 phase). As we introduce the disorder potential in the system, we observe the presence of the modes around quasienergy zero in the disorder-averaged eigenvalue spectrum as illustrated in Fig.~\ref{Eigenvalue}(a). The corresponding disorder-averaged local density of states~(LDOS) featuring corner localized modes is depicted in Fig.~\ref{Eigenvalue}(b). Interestingly, we also demonstrate the generation of disorder-induced $\pi$-modes in Fig.~\ref{Eigenvalue}(c) choosing one point within the clean R2 regime (denoted by the red star in Fig.~\ref{drive}). One can observe the presence of four $\pi$-modes from the disorder-averaged quasienergy spectrum [see Fig.~\ref{Eigenvalue}(c)]. The LDOS associated with the $\pi$-modes is depicted in Fig.~\ref{Eigenvalue}(d). Note that, the eigenvalue spectrum depicted in the inset of Fig.~\ref{Eigenvalue}(a) and Fig.~\ref{Eigenvalue}(c) indicates the absence of any topological modes around quasienergies $E_m=0/\pm\pi$ when disorder strength $V_{ij}=0$ in the trivial phase (R2). Thus, we have successfully demonstrated the generation of the disorder-induced FSOTAI phase, and this serves as the first prime result of this manuscript. Here, we discuss the spectral properties of the Floquet operator. In Appendix~\ref{App:LSSP}, we discuss the localization properties of the eigenstates of the Floquet operator in presence of disorder.

\section{Topological characterization} \label{SecIII}
To investigate the topological nature of the FSOTAI phase and find a phase diagram in the parameter space, we develop a topological invariant that can be employed to characterize the phase in the presence of disorder and most importantly should be able to distinguish between $0$- and $\pi$-modes. We exploit the chiral symmetry present in our system even in the presence of disorder. 
This symmetry demands the following relation in the presence of a periodic drive: $S H(t) S= - H(T-t)$, such that the time-evolution operator and the Floquet operator transform as $S U(t,0) S = U(T-t,0) 
U^\dagger(T,0)$ and $S U(T,0) S =  U^\dagger(T,0)$, respectively. We present this proof in Appendix~\ref{App:FOchiral}. Furthermore, the chiral symmetry allows us to divide the full time-period $T$ into two parts: first part $U_c=U(T/2,0)$ and second part $S U_c^\dagger S=U(T,T/2)$, such that one can define two operators $U_1=S U_c^\dagger S U_c$ and $U_2=U_c S U_c^\dagger S$. Here, $U_1$ and $U_2$ represent the full period evolution operator starting from $t=0$ and $t=T/2$, respectively and both of them satisfy $SU_{1,2} S = U_{1,2}^\dagger$~\cite{AsbothPRB2014}. Consequently, we can define two effective Hamiltonians corresponding to these two Floquet operators as $H_{\rm eff}^{1,2}= \frac{i}{T} \ln U_{1,2}$ so that 
they preserve the chiral symmetry: $S H_{\rm eff}^{1,2} S=-H_{\rm eff}^{1,2}$. We employ the chiral basis $U_S$ in which the chiral symmetry operator is diagonal such that $U_S^\dagger S U_S={\rm diag} \left(1,1,\cdots,-1,-1,\cdots \right)$. In this basis, the effective Hamiltonians possess an anti-diagonal form as
\begin{align}
    U_{S}^\dagger H_{\rm eff}^{1,2} U_{S} = \tilde{H}^{1,2}_{\rm eff}= \begin{pmatrix}
        0 & h_{1,2} \\
        h_{1,2}^\dagger & 0
    \end{pmatrix} \ .
    \label{windingnoformula}
\end{align}
Then, one can employ singular value decomposition of $h_j$ such that $h_j=U_A^j \Sigma_j U_B^j$; with $j=1,2$ and $\Sigma_j$ contain the singular values. Here, $A$ and $B$ represent different subspaces with positive and negative values of $U_S^{\dagger} S U_S$, respectively. Here, $U_\sigma^j=\left( \psi^{j,\sigma}_1,\psi^{j,\sigma}_2, \psi^{j,\sigma}_3, \cdots, \psi^{j,\sigma}_{N_\sigma}  \right)$ and $\psi^{j,\sigma}_n$'s are the eigenstates defined on subspace $\sigma(=A,B)$. Hence, we can define two winding numbers corresponding to $h_j$ as $\nu_j=\nu_j[h_j]$, such that~\cite{Ryu_2010,LinPRB2021,BenalcazarPRL2022}
\begin{align}
    \nu_j[h_j]=\frac{1}{2 \pi i} {\rm Tr} \left[ \log \left( \chi_A \chi_B^{-1} \right) \right] \ .
\end{align}
Limited to one-dimension, one may identify $\chi_\sigma$ as a polarization operator projected onto the $\sigma^{\rm th}$ sector of the eigenstate in the occupied band such that $\chi_\sigma=U^{-1}_\sigma P^\sigma  U_\sigma$~\cite{LinPRB2021}; with $P^\sigma=\sum_{i,\alpha \in \sigma} c^\dagger _{i,\alpha} \exp \left[ -i \frac{2 \pi}{L} x\right] c_{i,\alpha} $ \ie the polarization operator defined on the  $A$ or $B$ subspace. 

Translating towards a higher-order phase in two dimensions, we consider $\chi_\sigma$ to be the projected quadrupolar operator such that $\chi_\sigma=U^{-1}_\sigma Q^\sigma  U_\sigma$~\cite{BenalcazarPRL2022} where the quadrupolar operator on the sector $A$ or $B$ is defined as 
$Q^\sigma=\sum_{i,j,\alpha \in \sigma} c^\dagger _{i,j,\alpha} \exp \left[ -i \frac{2 \pi}{L^2} xy \right] 
c_{i,j,\alpha}$. With this definition of winding numbers, we now find the correct combination of $\nu_j$'s 
to obtain the number of $0$-/$\pi$-modes per corner~\cite{AsbothPRB2014}. In particular, $\nu_j$'s provide us with the difference between the total number of modes present per corner at subspace $A$ and $B$ corresponding to $H_{\rm eff}^j$. Thus, we can represent $\nu_j$'s as
\begin{align}
    \nu_j=&\left( n_{A,0}^j +n_{A,\pi}^j \right) - \left( n_{B,0}^j +n_{B,\pi}^j \right) \ ,
\end{align}
where, $n_{\sigma,\epsilon}^j$ represents the number of modes per corner residing on subspace 
$\sigma$ at quasienergy $\epsilon=0,\pi$. We investigate the location of the $0$-/$\pi$-modes on the subspace $A/B$ from $U_1$ and $U_2$. To this end, we consider $\ket{\Psi_1}$ to be the corner state and an eigenstate of $U_1$ such that $U_1 \ket{\Psi_1} = e^{-i \epsilon} \ket{\Psi_1}$; with $\epsilon \in \left\{ 0, \pi \right\}$. The state $\ket{\Psi_1}$ is also an eigenstate of the chiral symmetry operator: $S \ket{\Psi_1} = e^{-i \eta } \ket{\Psi_1}$ with $\eta=0$ and $\pi$ corresponding to $A$ and $B$, respectively. Thus, the corner mode can occupy only one of the subspaces. Afterward, we consider a corner state $\ket{\Psi_2}$ which is an eigenstate of $U_2$: $U_2 \ket{\Psi_2}=e^{-i \epsilon} \ket{\Psi_2}$. We find that $\ket{\Psi_2}=U_c \ket{\Psi_1}$ and the operation of the chiral symmetry on $\ket{\Psi_2}$ reads as $S \ket{\Psi_2}= e^{-i (\eta -\epsilon) }\ket{\Psi_2}$. Thus, we can now investigate which subspace the corner states $\ket{\Psi_1}$ and $\ket{\Psi_2}$ occupy at different quasienergies. In particular, for the $0$-modes, $\ket{\Psi_1}$ and $\ket{\Psi_2}$ occupy the same subspace while for the $\pi$-mode, $\ket{\Psi_2}$ and $\ket{\Psi_1}$ are on the different suspace. We show this explicitly in Appendix~\ref{App:endstates}. Thus, one obtains the following relations: $n_{A,\pi}^2-n_{B,\pi}^1=n_{B,\pi}^2-n_{A,\pi}^1=0$ and $n_{A,0}^2-n_{A,0}^1=n_{B,0}^2-n_{B,0}^1=0$. The number of $0$ and $\pi$ modes per corner is the difference between the number of modes at each subspace \ie $n_\epsilon=n_{A,\epsilon}-n_{B,\epsilon}$ with $n_{\sigma,\epsilon}=n^1_{\sigma,\epsilon}+n^2_{\sigma,\epsilon}$. Thus, we can define the topological index as the number of $0$ and $\pi$ modes per corner as 
\begin{align}
    W_0 = \frac{\nu_1+ \nu_2}{2}, \  {\rm and} \ W_\pi = \frac{\nu_1- \nu_2}{2}\ .
\end{align}
The above equation is the second prime result of this manuscript. We employ these winding numbers to topologically characterize the FSOTAI phase and chart out the phase diagram in the parameter space. 
This is depicted in Fig.~\ref{CleanPhase} in terms of $W_0$ and $W_\pi$. In particular, we obtain $(W_0,W_\pi)=(1,0)$ for R1, $(W_0,W_\pi)=(0,0)$ for R2, $(W_0,W_\pi)=(0,1)$ for R3, and 
$(W_0,W_\pi)=(1,1)$ for R4.

\begin{figure}
    \centering
    \subfigure{\includegraphics[width=0.49\textwidth]{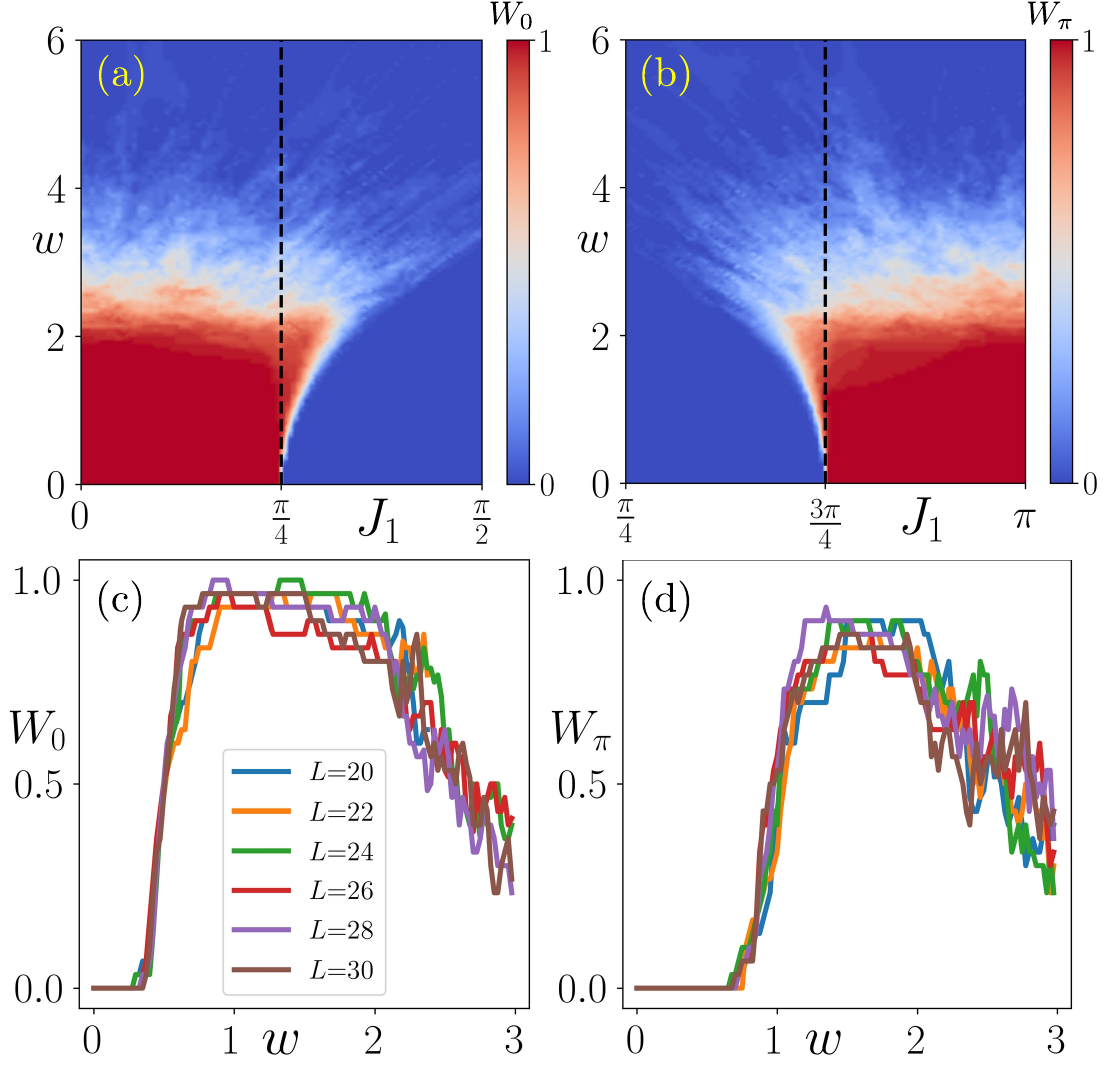}}
    \caption{We depict the phase diagram in the $J_1 \mhyphen w$ plane in terms of $W_0$ and $W_\pi$ in panels (a) and (b), respectively. The vertical black dashed line represents the topological region in the clean limit, while the extended region to the ``right and left" of that line indicate the FSOTAI phase as red color persists for $J_1> \pi/4$ and $J_1< 3\pi/4$ in panels (a) and (b), respectively. Our system comprises of $20 \times 20$ lattice sites and we take average over $50$ random disorder configurations. In panels (c) and (d), we show disorder averaged $W_0$ and $W_\pi$ as a function of disorder strength $w$ for different lattice sizes $L$ to showcase the disorder-driven topological transition. We choose $J_2=\pi/8$ and $T=2.0$ for all panels.
 }
    \label{DisorderPhase}
\end{figure}

For deeper understanding of the main results of this manuscript, we investigate the phase diagram of the FSOTAI in terms of $W_0$ and $W_\pi$ in $J_1 \mhyphen w$ plane for a fixed $J_2$ in Figs.~\ref{DisorderPhase}(a) and (b), respectively. The observation from these figures is two-fold- the stability of the $0$-/$\pi$-corner modes against onsite disorder strength $w$ and the disorder-driven topological phase transition. We notice that both the $0$-/$\pi$-modes are robust against the strong disorder strength $w\sim 2.5$ (red regions). Here, the disorder strength mediates a 
gap-closing transition, and the system becomes topologically trivial. Afterward, we discuss the generation of FSOTAI. In Figs.~\ref{DisorderPhase}(a) and (b), we depict a horizontal black dashed line to indicate the topological to non-topological phase transition with respect to $J_1$ for a clean system ($w=0$). We observe that the red regions broaden beyond the right [left] of the black line in Fig.~\ref{DisorderPhase}(a) [Fig.~\ref{DisorderPhase}(b)], indicating a non-zero value of $W_0$ [$W_\pi$]. This extended region represents a disorder-driven FSOTAI phase, which would be trivial in the absence of disorder. In Appendix~\ref{App:Windingstability}, we discuss the stability of the winding numbers $W_{0,\pi}$ in terms of variance corresponding to Figs.~\ref{DisorderPhase}(a) and (b). On the other hand, this extended topological region in Figs.~\ref{DisorderPhase}(a) and (b) is a consequence of the modification of the mass term originated due to the real part of the self-energy obtained from self-consistent Born's approximation~\cite{GrothPRL2009,LiPRL2020,ybyangPRB2021}. To emphasize more on this disorder-driven topological phase transition, we demonstrate $W_0$ and $W_\pi$ as a function of the disorder strength $w$ in Figs.~\ref{DisorderPhase}(c) and (d), respectively, for different lattice sizes $L$; with $L$ being the number of lattice sites in one direction. Note that, both $W_0$ and $W_\pi$ exhibit zero value for $w=0$. However, a non-zero value of $W_{0,\pi}$($\sim$1 for larger system size) is observed when we incorporate finite disorder strength ($w\neq 0$) in the system. Thus, the system exhibits corner modes only in the presence of disorder and periodic drive manifesting the FSOTAI phase. Here, we also mention that both $W_0$ and $W_\pi$ in Figs.~\ref{DisorderPhase}(c) and (d) are not exactly quantized, and this can be attributed to the finite size of the lattice. Nevertheless, we observe a finite region of $w$ depicting an FSOTAI phase.

In this section, overall we demonstrate the emergence of FSOTAI phase employing a three-step drive protocol. In Appendix~\ref{App:Harmonicdrive}, we supplement our results by employing a harmonic drive. Moreover, we also illustrate that our formalism works well for a different model, namely the Benalcazar-Bernevig-Hughes (BBH) model~\cite{benalcazar2017,benalcazarprb2017,Huang2020}.

\section{Summary and Discussion.} \label{SecIV}
To summarize, in this manuscript, we demonstrate the generation of FSOTAI phase (without any clean and static analog) hosting corner localized modes. We emphasize that both the disorder and periodic drive are key ingredients for the generation of this phase. Most importantly, we showcase the generation of $\pi$-modes in the FSOTAI phase, which is one of the primary results of this work. Moreover, we develop the real space topological invariant to characterize the driven-disordered phase. Our topological invariant only depends upon the presence of chiral symmetry, while it can be employed in the model that breaks mirror symmetries. Also, our formalism works for different driving protocols such as sinusoidal drive and other models (see Appendices~\ref{App:Harmonicdrive} and \ref{App:BBHmodelsec}). One can also formally generalize our developed topological invariant to characterize higher-order phases in higher dimensions, such as third-order topological phase in three dimensions, by identifying $\chi_\sigma$ as octupole operator in Eq.~(\ref{windingnoformula}). 

In recent times, HOTI has been demonstrated in different experimental setups~\cite{schindler2018higher,Experiment3DHOTI.VanDerWaals,Aggarwal2021,ShumiyaHOTI2022,Soldini2023,serra2018observation,xue2019acoustic,ni2019observation,Experiment3DHOTI.aSonicCrystals,
Ni2020,imhof2018topolectrical,PhotonicChen,PhotonicXie,mittal2019photonic}. Although, a second-order TI in two dimension is yet to be explored in a real material platform. Nevertheless, photonic systems can be the possible potential playground to observe the proposed FHOTAI phase. It has been shown that one can obtain photonic HOTI based on a silicon ring resonator 
in which the site-resonators act as the lattice sites while the link-resonators are used to generate the hopping. Then by adjusting the gap between the link- and the site-resonator, one can control the amplitude while by vertically or horizontally shifting the resonator one can change the sign of the coupling~\cite{mittal2019photonic}. 
By introducing two counter-propagating modes in each resonator, one can mimic the spins and pseudo-spin degrees of freedom; while by using semi-transparent scatters inside the site- or link-resonator, one can mix different degrees of freedom~\cite{Hafezi2011,LiangPRL2013}. The mismatch between the frequencies in a site-resonator introduces onsite disorder in the system. Moreover, it has been shown that these resonator-based systems can be modeled as networks, and the Bloch modes in these periodic network models can be mapped to the Floquet states~\cite{PasekPRB2014}. Thus, one can have all the building blocks to realize the Hamiltonians and the drive protocol [Eq.~(\ref{drive})]. Furthermore, one may also obtain the FHOTAI in an evanescently coupled photonic waveguides system, in which the propagation directions act as time and have already been reported to exhibit the Floquet TI phase hosting anomalous edge states~\cite{Maczewsky2017}. However, the main idea of this work has never been based on any specific system. We discuss the photonic system as a possible example which can mimic the similar physics. Our idea should be equally applicable to both 
solid state and cold atomic/photonic systems.
Therefore, given the rapid developments in sophisticated experimental techniques, we believe that our work paves the way for the experimental realization of the FSOTAI phase in the near future.


\begin{acknowledgments}
A.K.G. and A.S. acknowledge SAMKHYA: High-Performance Computing Facility provided by Institute of Physics, Bhubaneswar, for numerical computations. A.S. acknowledges Saikat Banerjee for useful comments and careful reading of the manuscript. 
\end{acknowledgments}


\appendix
\newcounter{defcounter}
\setcounter{defcounter}{0}

\section{Quasienergy spectra of clean FHOTI} \label{App:cleanFHOTI}
\begin{figure}
	\centering
	\subfigure{\includegraphics[width=0.49\textwidth]{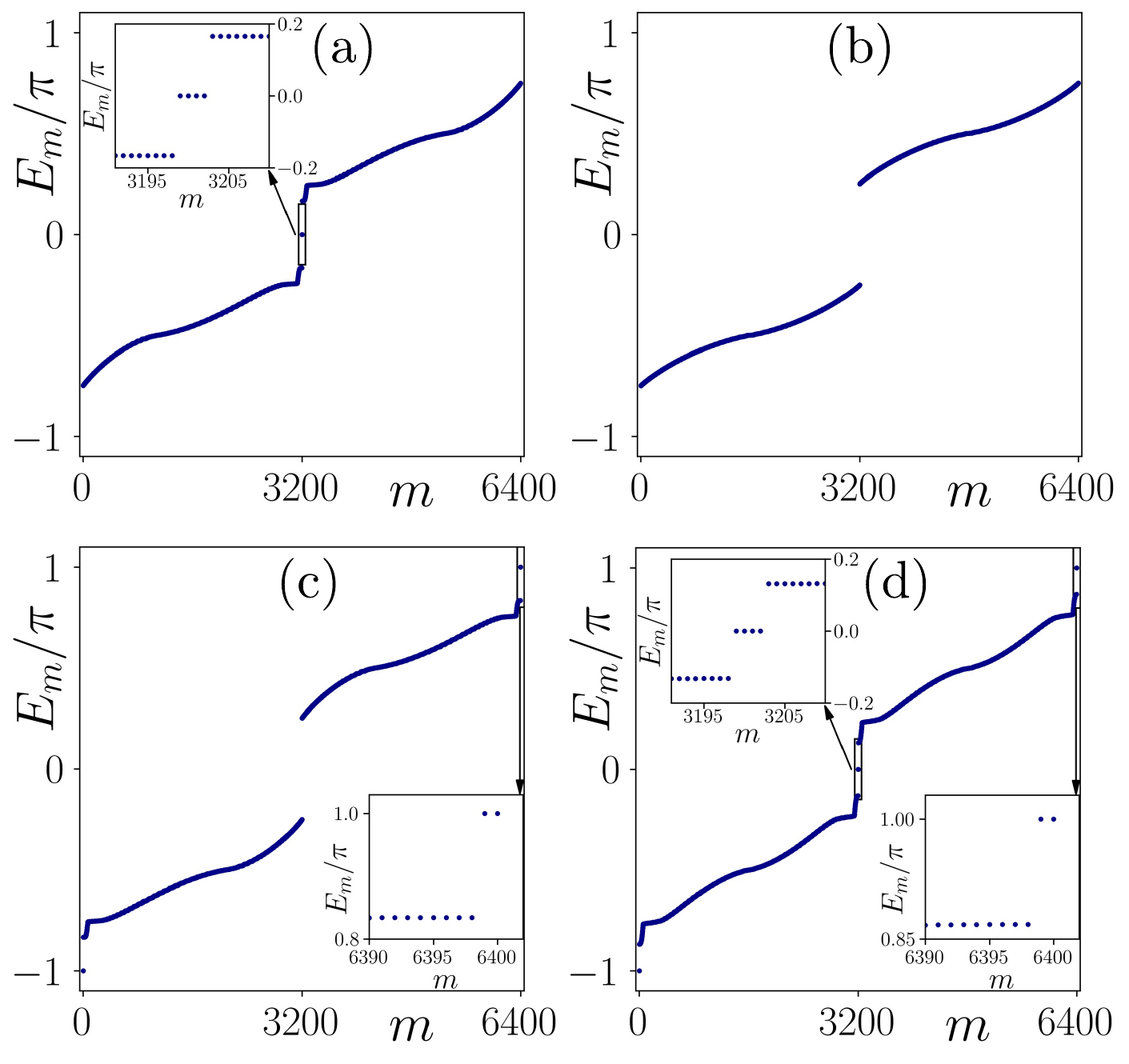}}
	\caption{In panels (a), (b), (c), and (d), we illustrate eigenvalue spectra $E_m$ as a function of the state index $m$ corresponding to R1, R2, R3, and R4, respectively. In the inset, we depict the topological states close to quasienergy $0$ or $\pi$ for better clarity. We choose the other model parameters as: $(J_1T,2 J_2T)=\left[ \left(\frac{\pi}{4},\frac{\pi}{2}\right),~\left(\frac{\pi}{2},\frac{\pi}{4}\right),~\left(\frac{3\pi}{4},\frac{\pi}{2}\right), ~\left(\frac{\pi}{2},\frac{3 \pi}{4}\right) \right]$ for R1, R2, R3, and R4, respectively. We consider a finite-size system consisting of $40 \times 40$ lattice sites.
	}
	\label{EigenvalueClean}
\end{figure}

In the absence of disorder, one can employ momentum space formalism and construct the Floquet operator $U(\vect{k};T,0)$ analytically~\cite{Huang2020,ghosh2021systematic,GhoshDynamical2022}. Afterward, employing the gap-closing condition at $\vect{k}=(0,0)/(\pi,\pi)$ for quasienergies $0$ and $\pi$, we can obtain the gap-closing relation as~\cite{Huang2020,ghosh2021systematic,GhoshDynamical2022}
\begin{align}
    \lvert J_2 \rvert T = \frac{\lvert J_1 \rvert T}{2} + m \pi \ ,
    \label{phaserel}
\end{align}
where, $m \in \mathbb{Z}$. We use Eq.~(\ref{phaserel}) to obtain the phase diagram of a clean FHOTI given our driving protocol. This is depicted in Fig.~\ref{CleanPhase}. Here, we also show the eigenvalue spectra of the system obeying open boundary condition (OBC) in Fig.~\ref{EigenvalueClean} when the driving parameters lie on the different parts of the phase diagram. In particular, we observe the presence of only $0$-mode, no modes (trivial), only  $\pi$-modes, and both $0$- and $\pi$-modes in Figs.~\ref{EigenvalueClean}(a), (b), (c), and (d), respectively. These correspond to R1, R2, R3, and R4, as depicted in Fig.~\ref{CleanPhase}.

\section{Localization properties}\label{App:LSSP}
\begin{figure}
	\centering
	\subfigure{\includegraphics[width=0.49\textwidth]{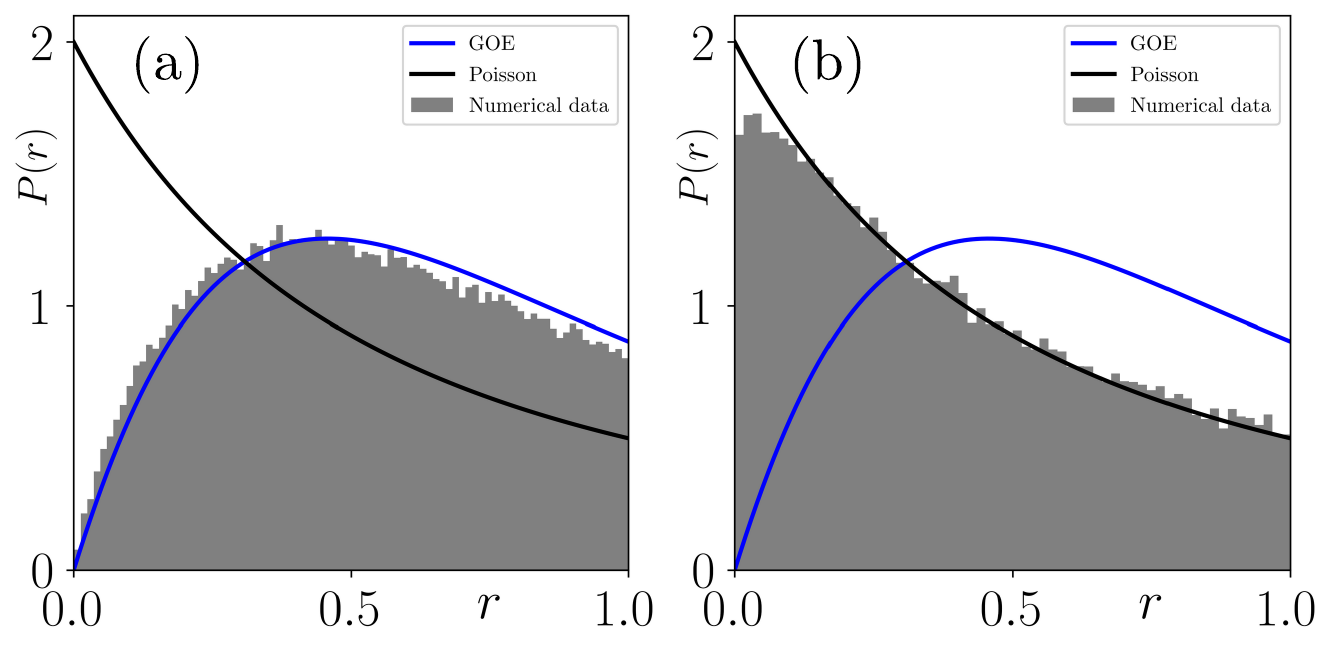}}
	\caption{In panels (a) and (b), we depict the probability distribution $P(r)$ as a function of the gap-ratio $r$ for disorder strength $w=0.5$ and $w=10$, respectively. The grey bars represent numerically obtained probability distributions. While the blue and black lines correspond to GOE and Poisson statistics, respectively. We consider a system consisting of $30 \times 30$ lattice sites with $100$ random disorder configurations. The rest of the system parameters take the same value as mentioned in Fig.~\ref{Eigenvalue}(a).
	}
	\label{LSS}
\end{figure}

Here, we discuss the localization properties of the eigenstates of the Floquet operator for our system. It is well known that the extended states exhibit level repulsion and obey Wigner-Dyson statistics while the localized states exhibit Poisson statistics~\cite{mehta2004random,OganesyanPRB2007,AlessioPRX2014,MarcoPRR2022}. To investigate the distributions of the eigenstates \ie the level spacing statistics~(LSS), we consider the gap between two consecutive states $\delta_n=E_{n+1}-E_{n}$, while the quasienergies $E_n$'s are sorted in the ascending order such that $\delta_n \ge 0$. We define the dimensionless quantity, gap ratio as $r_n={\rm min}\{\delta_n,\delta_{n-1}\}/{\rm max}\{\delta_n,\delta_{n-1}\}$~\cite{OganesyanPRB2007,AlessioPRX2014}. We depict the probability of the gap ratio in Fig.~\ref{LSS}. For small disorder strength ($w=0.5$), we find that the numerically obtained probability distributions (grey bars) match with Gaussian orthogonal ensemble (GOE) distributions $P_{\rm GOE}(r)=\frac{27}{4} \frac{r+r^2}{(1+r+r^2)^{5/2}}$ [see Fig.~\ref{LSS}(a)]. On the other hand, for strong disorder strength ($w=10$), we observe that the probability distribution exhibits Poissonian type distribution  $P_{\rm Poisson}(r)=\frac{2}{(1+r)^2}$ [see Fig.~\ref{LSS}(b)]. Thus, for a small (strong) disorder, our system exhibits extended (localized) states. Hence, the localization properties we discuss here are owing to the bulk states only, manifesting delocalization to localization transition as the disorder strength is increased. However, it is still an open question how does LSS differentiate FSOTAI phase from regular Anderson insulator phase.

\section{Floquet operator under chiral symmetry} \label{App:FOchiral}
Here, we discuss how the time evolution operator $U(t,0)$ and the Floquet operator $U(T,0)$ transform under chiral symmetry. The time periodic Hamiltonian $H(t)$ transforms under chiral symmetry
as $SH(t)S=-H(T-t)$. Thus, one can obtain
\begin{align}
    S  U & (t,0) S \non \\
    =& S \left[  {\rm TO} \exp \left( -i \int_0^t H(t') dt' \right) \right] S \non \\
    =& \sum_m \frac{(-i)^m}{m!} {\rm TO} \int_0^t dt_1' \  \cdots \ \int_0^t dt_m' SH(t_1')S \ \cdots \ SH(t_m')S \non \\
    =& \sum_m \frac{(i)^m}{m!} {\rm TO} \int_0^t dt_1' \  \cdots \ \int_0^t dt_m' H(T-t_1') \ \cdots \ H(T-t_m') \non \\
    =& \sum_m \frac{(-i)^m}{m!} {\rm TO} \int_T^{T-t} dt_1' \  \cdots \ \int_T^{T-t} dt_m' H(t_1') \ \cdots \ H(t_m') \non \\
    =& {\rm TO} \exp \left( -i \int_T^{T-t} H(t') dt' \right) \non \\
    =& U(T-t,T) \non \\
    =& U(T-t,0) U(0,T) \non \\
    =& U(T-t,0) U^\dagger(T,0) \ .
\end{align}
Thus, the Floquet operator transform as $ S U(T,0) S = U^\dagger(T,0) $.

\begin{table*}[!ht]
\centering
\begin{tabular}{cl|c|c|ll}
\cline{3-4}
\multicolumn{1}{l}{}                                      &                              & $\ket{\Psi_1}$: $S \ket{\Psi_1}= e^{-i (\gamma- \epsilon)}\ket{\Psi_1}$ & $\ket{\Psi_2}$: $S \ket{\Psi_2}= e^{-i (\gamma- \epsilon)}\ket{\Psi_2}$ &  &  \\ \cline{1-4}
\multicolumn{1}{|c|}{\multirow{2}{*}{For $\epsilon=0$}}   & Subspace $A$: $\gamma=0$   & $S \ket{\Psi_1}=\ket{\Psi_1}$                                           & $S \ket{\Psi_2}=\ket{\Psi_2}$                                           &  &  \\ \cline{2-4}
\multicolumn{1}{|c|}{}                                    & Subspace $B$: $\gamma=\pi$ & $S \ket{\Psi_1}=-\ket{\Psi_1}$                                          & $S \ket{\Psi_2}=-\ket{\Psi_2}$                                          &  &  \\ \cline{1-4}
\multicolumn{1}{|c|}{\multirow{2}{*}{For $\epsilon=\pi$}} & Subspace $A$: $\gamma=0$   & $S \ket{\Psi_1}=\ket{\Psi_1}$                                           & $S \ket{\Psi_2}=-\ket{\Psi_2}$                                          &  &  \\ \cline{2-4}
\multicolumn{1}{|c|}{}                                    & Subspace $B$: $\gamma=\pi$ & $S \ket{\Psi_1}=-\ket{\Psi_1}$                                          & $S \ket{\Psi_2}=\ket{\Psi_2}$                                           &  &  \\ \cline{1-4}
\end{tabular}
\caption{Transformation of $\ket{\Psi_1}$ and $\ket{\Psi_2}$ under chiral symmetry $S$.}
\label{tableChiral}
\end{table*}
\section{Investigation of the corner modes under chiral symmetry}\label{App:endstates}
Here, we investigate the transformation of the end states $\ket{\Psi_1}$ and $\ket{\Psi_2}$ under chiral symmetry $S$. This is presented in the Table~\ref{tableChiral}. By inspecting Table~\ref{tableChiral}, we can conclude that $\ket{\Psi_2}$ occupies the same and opposite subspace for zero- and $\pi$-modes, respectively.

\begin{figure}
	\centering
	\subfigure{\includegraphics[width=0.49\textwidth]{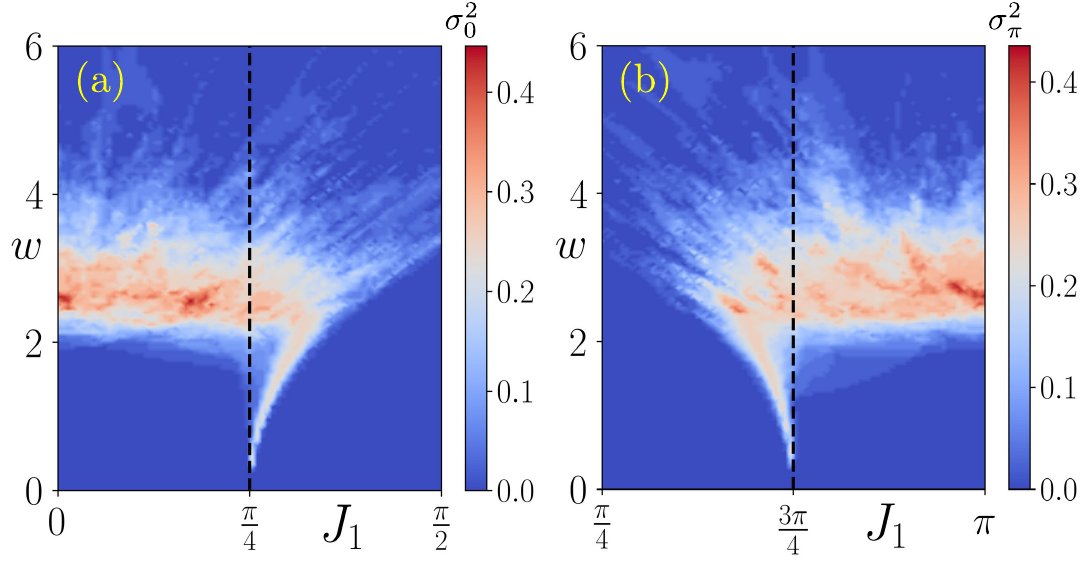}}
	\caption{In panels (a) and (b), we illustrate $\sigma_0^2$ and $\sigma_\pi^2$, respectively, in the $J_1 \mhyphen w$ plane. The model parameters take the same value as mentioned in Fig.~\ref{DisorderPhase}.
	}
	\label{Variance}
\end{figure}

\section{Investigation of the stability of the Winding numbers}\label{App:Windingstability}
To investigate the stability of the winding numbers ($W_{\epsilon}$) employed to obtain the phase diagram discussed in Fig.~3 (main text), we investigate the variance $\sigma_\epsilon^2$ 
of $W_\epsilon$, defined as
\begin{align}
    \sigma_\epsilon^2=\frac{1}{N}\sum_{i}^{N} \left(W_{\epsilon,i} - W_{\epsilon} \right)^2\ ,
\end{align}
where, $\epsilon=0,\pi$. Here $N$, $W_{\epsilon,i}$, and $W_{\epsilon}$ denote the total number of disorder configurations, the winding number for the $i^{\rm th}$ disorder configuration, and the average winding number, respectively. In Fig.~\ref{Variance}(a) and (b), we depict the variance $\sigma_0^2$ and $\sigma_\pi^2$ in the $J_1 \mhyphen w$ plane corresponding to $0$- and $\pi$-modes, respectively. We find that the variance becomes smaller in magnitude (blue to light blue) for a finite range of disorder strength $w$ and especially for the FSOTAI phase highlighting the robustness 
of this phase.

\section{Harmonically driven disordered setup}\label{App:Harmonicdrive}
In the main text, we discuss the generation of FSOTAI employing step drive protocol. Here, we introduce a sinusoidal driving protocol to showcase the applicability of the winding numbers. To start with,
we consider the following Hamiltonian, which is a combination of the Bernevig-Hughes-Zhang~(BHZ) model and a $C_4$ and TRS breaking Wilson-Dirac (WD) mass term~\cite{GhoshTimedynamics2023}:
\begin{align}
H_0=& \sum_{i,j} c_{i,j}^\dagger  \big[  (M-4B + V_{ij}) \Gamma_1 c_{i,j}+ B \Gamma_1 c_{i+1,j}+ B \Gamma_1  c_{i,j+1}  \non \\
&-  \frac{iA}{2} \Gamma_2 c_{i+1,j} - \frac{iA}{2} \Gamma_3 c_{i,j+1} + \frac{\Lambda}{2} \Gamma_4 c_{i+1,j} - \frac{\Lambda}{2} \Gamma_4 c_{i,j+1} \big] \non \\
&+ {\rm h.c.} ,\ 
\end{align}
where, $M$, $B$, $A$, and $\Lambda$ represent crystal field splitting (staggered chemical potential), hopping amplitude, spin-orbit coupling, and the strength of the WD mass term, respectively.
Here, $\Gamma_1=\sigma_z s_0$, $\Gamma_2=\sigma_x s_z$, $\Gamma_3=\sigma_y s_0$, $\Gamma_4=\sigma_x s_x$ and $V_{ij}$ denote the random onsite disorder potential such that $V_{ij} \in \left[ -\frac{w}{2},\frac{w}{2} \right]$. While the $4 \times 4$ $\vect{\Gamma}$ matrices are the same as defined in the main text. Then, we introduce the driving protocol in terms of sinusoidal variation of 
onsite mass term as
\begin{align}
H_1(t)= \sum_{i,j} c_{i,j}^\dagger  V_0 \cos \Omega t \Gamma_1 c_{i,j} \ ,
\end{align}
where, $V_0$ and $\Omega$ are the driving strength and the driving frequency, respectively. The full time-dependent Hamiltonian $H(t)=H_0+H_1(t)$ is periodic in time such that $H(t+T)=H(t)$. 
We can construct the time evolution operator as~\cite{GhoshTimedynamics2023}
\begin{align}
    U(t,0)=& {\rm TO} ~ \exp \left[ -i \int_{0}^{t} dt' H(t') \right]  \non \\
    =& \prod_{j=0}^{N-1} U(t_{j}+\delta t,t_j) \ ,
\end{align}
where, $U(t_{j}+\delta t,t_j)=e^{-i {H}(t_j)\delta t}$; with $\delta t = \frac{t}{N_t}$, $t_j=j \delta t$, and $N_t$ is the number of time-steps. Following this evolution operator, we can follow the procedure discussed in the main text and compute the winding numbers $W_0$ and $W_\pi$. 
\vspace{0.15cm}
\begin{figure}
	\centering
	\subfigure{\includegraphics[width=0.49\textwidth]{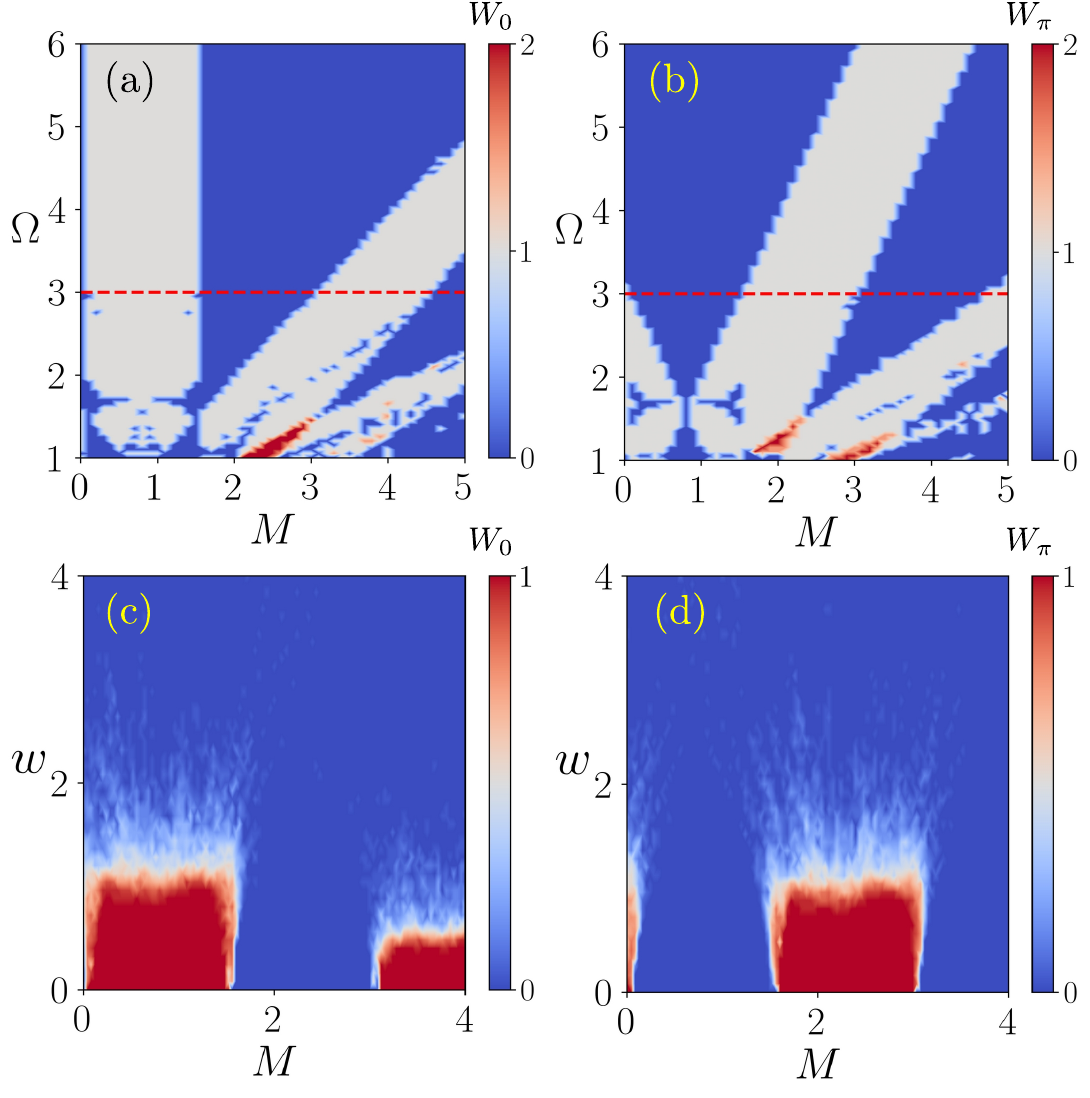}}
	\caption{In panels (a) and (b), we illustrate the phase diagram in the $M \mhyphen \Omega$ plane for the $0$-mode and $\pi$-mode, respectively, considering the sinusoidal drive. 
	The color bars indicate the magnitudes of $W_0$ and $W_\pi$. We choose $A=B=\Lambda=0.2$ and $V=2.0$. We depict the phase diagram in terms of $W_0$ and $W_\pi$ in the 
	$M \mhyphen w$ plane in panels (c) and (d), respectively. We choose the drive frequency $\Omega=3.0$ as indicated by the red dashed line in panels (a) and (b).
	}
	\label{Harmonic}
\end{figure}

First, we investigate the phase diagram for a clean-periodically driven system \ie $w=0$. We depict $W_0$ and $W_\pi$ in the $M \mhyphen \Omega$ plane for the disorder-free case in Figs.~\ref{Harmonic}(a) and (b), respectively. We observe that at the low-frequency regime, we obtain topological phases where both $W_0$ and  $W_\pi$ exhibit a value of two, which is an indication of the generation of multiple (two) modes per corner. Thus, the winding numbers can be employed to obtain the number of states per corner, which is another advantage compared to that of the nested-Wilson loop technique~\cite{Huang2020,GhoshDynamical2022}. 
After discussing the phase diagram of a clean system, we move our attention towards the disordered case. We demonstrate the phase diagram for the $0$- and $\pi$-modes in Figs.~\ref{Harmonic}(c) 
and (d), respectively. We observe that the corner states are robust against the disorder up to a certain disorder strength $w$. However, we do not obtain any clear indication of the generation of Floquet topological Anderson phase that we observe clearly for the step drive case [see Fig.~\ref{DisorderPhase}]. The reason can be attributed to the fact that we consider a relatively small system size ($16 \times 16$), as the computation for the Harmonic drive is numerically costly compared to the step drive. Thus, for the Harmonic drive, one may observe the FSOTAI phase with a larger system size. Nevertheless, the aim of the current manuscript is to demonstrate the generation of FSOTAI in a driven-disordered setup, which one can accomplish via step drive as well. 

\begin{figure}
	\centering
	\subfigure{\includegraphics[width=0.49\textwidth]{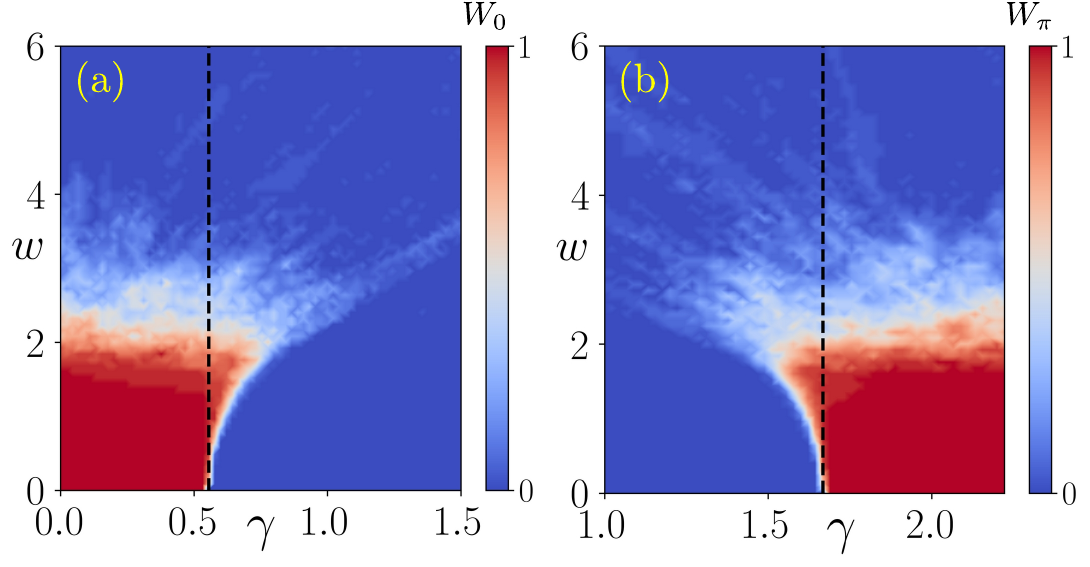}}
	\caption{In panels (a) and (b), we depict $W_0$ and $W_\pi$ in the $\gamma-w$ plane respectively, for a driven BBH model. We choose the other model parameters as $\lambda=\pi/4\sqrt{2}$ 
	and $T=2$.
	}
	\label{BBHmodel}
\end{figure}
\section{Driven BBH model}\label{App:BBHmodelsec}
In the main text, we showcase the generation of the FSOTAI phase employing mirror symmetry-breaking BHZ model and a WD mass term. Here, we consider a mirror symmetry preserving model, in particular, a driven BBH model~\cite{benalcazar2017,benalcazarprb2017,Huang2020}. We consider the same driving protocol as introduced by Eq.~\eqref{drive}, 
given as
\begin{align}
H (t)&=  h_1 \  ;   \quad \quad t \in [0, T/4] \ , \nonumber \\
&= h_2 \  ; \quad \quad t \in (T/4,3T/4] \  , \nonumber \\
&= h_3 \  ; \quad \quad t \in (3T/4,T ] \ ,
\label{driveBBH}
\end{align}
where, the step Hamiltonians $h_1$ and $h_2$ in the real space reads as
\begin{align}
    h_1=&h_3= \sum_{i,j} c_{i,j}^\dagger \left( \gamma + V_{ij} \right) (\tilde{\Gamma}_1-\tilde{\Gamma}_2) c_{i,j} \ , \non  \\
    h_2=& \sum_{i,j} c_{i,j}^\dagger \frac{\lambda}{2} \big( \tilde{\Gamma}_1  c_{i+1,j}+ \tilde{\Gamma}_2  c_{i,j+1}  - i \tilde{\Gamma}_3 c_{i+1,j} - i \tilde{\Gamma}_4 c_{i,j+1} \big) \non \\
    &\quad + {\rm h.c.} \ ,
\end{align}
with $\tilde{\Gamma}_1=\sigma_x s_0$, $\tilde{\Gamma}_2=\sigma_y s_y$, $\tilde{\Gamma}_3=\sigma_y s_z$, and $\tilde{\Gamma}_4=\sigma_y s_x$. Here, $\vect{\sigma}$ and $\vect{s}$ operate on two different types of orbital degrees of freedom. The driven system is controlled via $\gamma$ and $\lambda$. The disorder potential $V_{ij}$ is randomly distributed as $V_{ij} \in \left[ -\frac{w}{2},\frac{w}{2} \right]$. The phase diagram for this system resembles that of shown in Fig. 1 in the main text. Thus, we choose two points from region R2 (trivial phase) and investigate the generation of FSOTAI. 
We illustrate the phase diagram in terms of $W_0$ and $W_\pi$ in the $\gamma \mhyphen w$ plane in Figs.~\ref{BBHmodel}(a) and (b), respectively. The vertical black dashed line represents the topological to non-topological phase transition point for the clean system. In the presence of disorder, one can note that there is an extension of the topological phase beyond the topological regime 
in the clean system and this extended regime indicate the FSOTAI phase hosting corner localized modes.

\bibliographystyle{apsrev4-2mod}
\bibliography{bibfile.bib}
\end{document}